\documentclass[12pt,epsfig]{article}
\usepackage{graphicx,amsmath}

\newcommand{\be}{\begin{equation}}
\newcommand{\ee}{\end{equation}}
\newcommand{\bea}{\begin{eqnarray}}
\newcommand{\eea}{\end{eqnarray}}

\newcommand{\p}{\vec{p}}

\newcommand{\func}{\mbox}

\begin{document}

\begin{titlepage}

\begin{flushright}
\begin{tabular}{l}
 CPHT RR 060.0602
\end{tabular}
\end{flushright}
\vspace{1.5cm}

\begin{center}

{\LARGE \bf  Pomeron-Odderon interference effects in
\vskip.1in
 electroproduction of two pions}
\vspace{1cm}

{\sc Ph.~H{\"a}gler}${}^{a}$,
{\sc B.~Pire}${}^{b}$,
{\sc L.~Szymanowski}${}^{b,c}$ and
{\sc O.V.~Teryaev}${}^{d}$
\\[0.5cm]
\vspace*{0.1cm} ${}^a${\it
   Institut f{\"u}r Theoretische Physik, Universit{\"a}t
   Regensburg, \\ D-93040 Regensburg, Germany
                       } \\[0.2cm]
\vspace*{0.1cm} ${}^b$ {\it
CPhT, {\'E}cole Polytechnique, F-91128 Palaiseau, France\footnote{
  Unit{\'e} mixte C7644 du CNRS.}
                       } \\[0.2cm]
\vspace*{0.1cm} ${}^c$ {\it
 So{\l}tan Institute for Nuclear Studies,
Ho\.za 69,\\ 00-681 Warsaw, Poland
                       } \\[0.2cm]
\vspace*{0.1cm} ${}^d$ {\it
Bogoliubov Lab. of Theoretical Physics, JINR, 141980 Dubna, Russia
                       } \\[1.0cm]
{\it \large
\today
 }
\vskip2cm
{\bf Abstract:\\[10pt]} \parbox[t]{\textwidth}{
We study Pomeron-Odderon interference effects giving rise to 
charge and single-spin
asymmetries in diffractive electroproduction of a
 $\pi^+\;\pi^-$ pair. We calculate these asymmetries originating from
both longitudinal and transverse polarizations of the virtual photon, in
the
framework of
QCD and in the Born approximation, in a kinematical domain accessible to
HERA
experiments. We predict a sizable charge asymmetry with a 
characteristic dependence on the invariant mass of the $\pi^+\,\pi^-$ pair,
which makes
this observable very important for establishing the magnitude of 
the Odderon exchange
in hard processes. The single spin asymmetry turns out to be rather small.
We briefly discuss future improvements of our calculations and their 
possible effects on the results. 

 }
\vskip1cm
\end{center}

\vspace*{1cm}

\end{titlepage}

\section{Introduction}


Hadronic reactions at low momentum transfer and high energies are described
 in the framework of QCD in terms of the dominance of color singlet
exchanges
corresponding to a few reggeized gluons. 
The charge conjugation even 
sector of the $t-$channel exchanges is
understood as the QCD-Pomeron described by the BFKL equation \cite{BFKL}. The
charge-odd exchange is less well understood although the corresponding 
BKP equations \cite{BKP} have attracted much attention
recently \cite{Levodd,JW,Vacca1,Korch}, thus
reviving the relevance of phenomenological studies of the Odderon exchange
pointed
out years ago in Ref. \cite{LN}. Recent studies \cite{Doshrecent} confirm
indeed the
need for the Odderon contribution, in particular to understand the different
behaviours of $pp$ and $\bar p p$ elastic cross sections in the dip region.
However studies of specific channels where the Odderon contribution is expected
to be
singled out have turned out to be very disappointing. 
Recent experimental studies at HERA of
exclusive $\pi^0$ photoproduction  \cite{Olsson}  indicate a very
small cross section for this process which stays in contradiction with
theoretical
predictions based on the stochastic vacuum model \cite{Dosh}.
In  diffractive
$\eta_c$-meson
photoproduction,   the  QCD prediction for
the cross
section is rather small \cite{KM,Engel} at Born level;  the inclusion of
evolution
following from
the BKP equation
\cite{Vacca2} leads to an increase of the predicted
cross section for this process by  one order of magnitude but 
no experimental data exist so far.

\vskip.1in
A new strategy to reveal the features of the charge-odd exchange is thus
required. For that purpose let us first note that
in all above mentioned
meson production processes the scattering amplitude describing Odderon
exchange enters quadratically in the cross section.
This observation lead to the suggestion
in Ref. \cite{Brodsky}, that the study of observables where Odderon effects
are present  at the amplitude level - and not at the squared amplitude
level - is
mandatory to get a convenient sensitivity to a rather small normalization
of this contribution.
 This may be achieved by means of  charge asymmetries, as for
instance in open charm production \cite{Brodsky}. Since the final state
quark-antiquark pair has no definite charge parity both Pomeron and Odderon exchanges
contribute to this process. Another example \cite{Nikolaev} is the charge asymmetry in soft
photoproduction of two pions. 
On the other hand, the difficulty with
the understanding of soft processes in QCD calls for studies of Odderon
contributions in hard processes, such as electroproduction, where factorization
properties allow for a perturbative calculation of the  short-distance part of the
scattering amplitude.

\vskip.1in 

In a recent paper \cite{HPST} we proposed to 
  study
the diffractive electroproduction
of a $\pi^+ \,\pi^-$ pair 
to search for the QCD-Odderon at the amplitude level. 
The $\pi^+\,\pi^-$-state
doesn't have  any definite charge parity and therefore both Pomeron and
Odderon exchanges contribute. 
The originality of our study of the electroproduction
process with respect to Refs. \cite{Brodsky,Nikolaev} is to work in a perturbative QCD
framework
which enables us to
derive more founded predictions in an accessible kinematical domain.

\vskip.1in In this paper, we study   in full detail the
charge and single spin asymmetries in the deeply virtual 
 production of two
pions
within perturbative QCD, see Fig. 1. The application of pQCD for the calculation of a
part of this process is justified by the presence of a 
 hard scale: the squared mass $-Q^2$ of the virtual photon,
$Q^2$ being of the order of a few GeV$^2$.
The amplitude of this process
 includes the convolution of a
perturbatively calculable hard subprocess with two non-perturbative
inputs, 
the two pion generalized distribution amplitude (GDA) and the Pomeron-Odderon (P/O)
proton impact factors. Since
the $\pi^+\pi^-$ system is not a  charge parity
eigenstate, the GDA includes two charge parity components and allows for
a study of the corresponding interference term. The relevant GDA is here
just given by the light cone wave function of the two pion system \cite{DGPT}.


In this paper we supplement our previous work \cite{HPST} by
the inclusion 
of contributions from transversely
polarized photons to the charge asymmetry. 
Additionally we study
 the single spin asymmetry which
is proportional to the interference of non-diagonal, i.e.
longitudinal-transverse-polarization, terms. In contrast the charge
asymmetry picks up contributions of all possible polarization
combinations, which, as expected,  turn out to be of the same order of
magnitude at
moderate $Q^2$.


\vskip.1in

Since transversely polarized pion pairs are the only source
of a dependence of 
the
amplitude  on the azimuthal angle of the pions in their c.m.
frame, the amplitudes and cross sections are independent of this angle in
our approximation. As a result, the transverse charge asymmetry
\cite{Nikolaev}, resulting from a distribution in this angle, is
zero.

\vskip.1in Our results for the charge and spin asymmetries, which have been
obtained by a lowest order calculation, can be extended by the inclusion
of evolution from the BFKL and BKP equations, in a similar way as it has
been done in  \cite{Vacca2}.


\section{Kinematics}

Let us first specify the kinematics of the process under study, namely the
electron proton scattering
\begin{equation}  \label{ep}
e(p_i, \lambda)\;\; N (p_N) \to e(p_f)\;\;\pi^+(p_+)\;\; \pi^-(p_-)\;\;
N^{\prime}(p_{N^{\prime}})\;.
\end{equation}
which proceeds through a virtual photon-proton reaction (Fig.~1)
\begin{equation}  \label{gp}
\gamma^* (q, \epsilon)\;\; N (p_N) \to \pi^+(p_+)\;\; \pi^-(p_-)\;\;
N^{\prime}(p_{N^{\prime}})\;.
\end{equation}
where $\lambda$ is the initial electron helicity and $\epsilon$ the virtual
photon polarization vector.


\begin{figure}[t]
\centerline{\includegraphics[scale=.5]{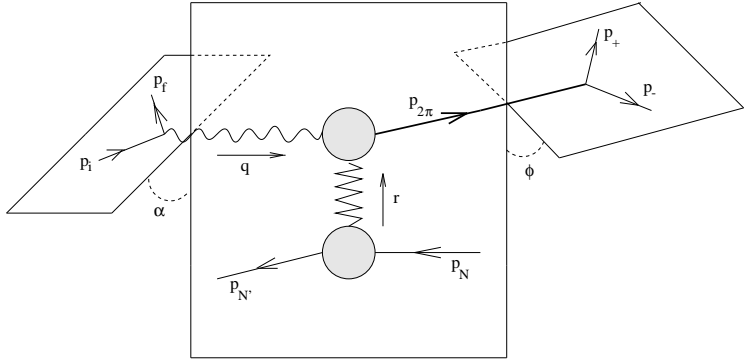}}
\caption{{\protect\small Kinematics of the electroproduction of two pions }}
\label{fig:1}
\end{figure}
%

We introduce a Sudakov representation of all particle momenta using the Sudakov
light-like momenta $p_1,p_2$. The virtual photon momentum can then be
written as
\begin{equation}
q^{\mu }=p_{1}^{\mu }-\frac{Q^{2}}{s}p_{2}^{\mu }
\label{q}
\end{equation}
where $s = 2p_1 \cdot p_2$. Similarly, 
the nucleon momentum in the initial state can be expressed through
\begin{equation}
p_{N}^{\mu} = p_2^{\mu} +\frac{M^2}{s}p_1^{\mu}\;,
\end{equation}
where $M$ is the proton target mass. The variable $s$
  is related to the total
energy squared of the virtual photon - proton system by
\[
(q + p_N)^2 \approx s - Q^2 +M^2\ \approx s\;.
\]
As usual, $y$ is the energy fraction carried by the virtual photon

\begin{equation}
\label{y}
y = \frac{q.p_2}{p_i.p_2}.
\end{equation}
The momentum of the two pion system is given by
\begin{equation}  \label{2pi}
p_{2\pi}^{\mu} = (1-\frac{\vec{p}_{2\pi}^{\;2}}{s})p_1^{\mu} +\frac{m_{2\pi}^2 +
\vec{p}_{2\pi}^{\;2} }{s}p_2^{\mu} + p_{2\pi \perp}^{\mu},\;\;\;\; p_{2\pi
\perp}^2=-\vec{p}_{2\pi}^{\;2}\ .
\end{equation}
We denote by $\alpha$ the angle between the euclidean vectors $\vec p_i$ and $\vec
p_{2\pi}$.

The quark momentum $l_1$ and antiquark momentum $l_2$ inside the loop before
the formation of  the two pion system (see Fig.~2) are parametrized as
\begin{equation}  \label{l1}
l_1^{\mu} = z p_1^{\mu} +\frac{m^2+(\vec{l}+z\vec{p}_{2\pi})^2}{zs}%
p_2^{\mu}+ (l_\perp +z p_{2\pi\ \perp})^{\mu}
\end{equation}
\begin{equation}  \label{l2}
l_2^{\mu} = \bar{z} p_1^{\mu} +\frac{m^2+(-\vec{l}+ \bar{z}\vec{p}_{2\pi})^2%
}{\bar{z}s}p_2^{\mu}+ (-l_\perp +\bar{z}\ p_{2\pi\ \perp})^{\mu}
\end{equation}
where $2\vec{l}$ is the relative transverse momentum of the quarks forming
the two pion system and $\bar{z} = 1-z $, up to small corrections of the
order $\vec{p}_{2\pi}^{\;2}/s$. Following the collinear approximation of the
factorization
procedure in the description of the two pion formation through the
generalized distribution amplitude we put 
$\vec{l}=\vec{0}$
in the hard amplitude.

In a similar way as in (\ref{l1}), (\ref{l2}) we parametrize the momenta of
the produced pions as
\begin{equation}  \label{p+}
p_+^{\mu} = \zeta p_1^{\mu} +\frac{m_\pi^2+(\vec{p}+ \zeta\vec{p}_{2\pi})^2}{%
\zeta s}p_2^{\mu}+ (p_\perp +\zeta p_{2\pi \perp})^{\mu}
\end{equation}
\begin{equation}  \label{p-}
p_-^{\mu} = \bar{\zeta} p_1^{\mu} +\frac{m_\pi^2+(-\vec{p}+ \bar{\zeta }\vec{%
p}_{2\pi})^2}{\bar{\zeta }s}p_2^{\mu}+ (-p_\perp +\bar{\zeta }\ p_{2\pi
\perp})^{\mu}
\end{equation}
where $2\vec{p}$ is now their relative transverse momentum, 
$\zeta = \frac{p_2\cdot p_+}{p_2\cdot p_{2\pi}}$ is the fraction of
the longitudinal momentum $p_{2\pi}$ carried by the produced $\pi^+$, and 
$\bar{\zeta } = 1-\zeta $.  The variable $\zeta$ is related to the polar
decay angle $\theta$ which is  in the rest frame of the pion pair
defined by
\begin{equation}  \label{theta}
\beta \cos \theta = 2\zeta -1\,,\;\;\; \beta \equiv \sqrt{1 - \frac{%
4\,m_\pi^2}{m_{2\pi}^2}}\;.
\end{equation}
Since the "longitudinal part" of the two pion wave function depends only on
the angle $\theta$ and doesn't depend on the azimuthal decay angle $\phi$
(in the same rest frame of the pair) we focus on the calculation of
forward-backward asymmetries expressed in terms of $\theta$ (see below).
The squared momentum transfer $t=r^2\ $($r^{\mu} = p_{2\pi}^{\mu} -q^{\mu})$
can be written as
\begin{equation}
t=r^2= -\vec{p}_{2\pi}^{\;2} + t_{min}, \;\;\;\;t_{min}= -\frac{%
M^2(Q^2+m_{2\pi}^2)^2} {s^2}\;.
\end{equation}

\section{Scattering amplitudes}

It is well known (see e.g. \cite{IF}, \cite{Engel} and references therein) that
for
large values of $s$, large $Q^2$ and small momentum transfer $t$ the scattering
amplitudes can be represented as convolutions over the two-dimensional
transverse momenta of the $t$-channel gluons.

\begin{figure}[t]
\centerline{\includegraphics[scale=.6]{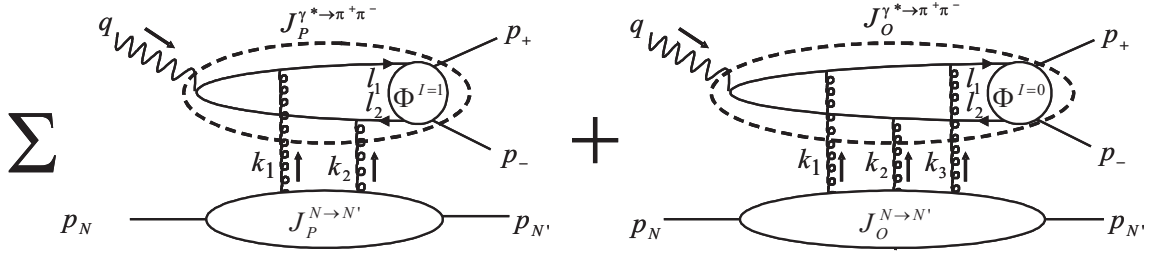}}
\caption{{\protect\small Feynman diagrams describing $\pi^+ \pi^-$
electroproduction in the Born approximation }}
\label{fig:2}
\end{figure}
%

For the Pomeron exchange, which corresponds in the Born approximation of
QCD to the
exchange of two gluons in a colour singlet state, see Fig. 2,  the impact
representation
has the form
\begin{equation}  \label{pom}
\mathcal{M}_P = -i\,s\,\int\;\frac{d^2 \vec{k}_1 \; d^2 \vec{k}_2 \;
\delta^{(2)}(\vec{k}_1 +\vec{k}_2-\vec{p}_{2\pi})}{(2\pi)^2\,\vec{k}_1^2\,%
\vec{k}_2^2} J_P^{\gamma^* \rightarrow \pi^+\pi^-}(\vec{k}_1,\vec{k}_2)\cdot
J_P^{N \rightarrow N^{\prime}}(\vec{k}_1,\vec{k}_2)
\end{equation}
where $J_P^{\gamma^* \rightarrow \pi^+\pi^-}(\vec{k}_1,\vec{k}_2)$  and $%
J_P^{N \rightarrow N^{\prime}}(\vec{k}_1,\vec{k}_2)$ are the impact factors
for the transition $\gamma^* \to \pi^+\ \pi^-$ via Pomeron exchange  and
of the
nucleon in the initial state $N$ into the nucleon in the  final state
$N^{\prime}$.

The corresponding representation for the Odderon exchange, \emph{%
i.e.} the exchange of three gluons in a colour singlet state, is given by
the formula
\begin{equation}  \label{odd}
\mathcal{M}_O =-\frac{8\,\pi^2\,s}{3!}\int\;\frac{d^2 \vec{k}_1 \; d^2 \vec{k%
}_2 d^2 \vec{k}_3\; \delta^{(2)}(\vec{k}_1 +\vec{k}_2 +\vec{k}_3-\vec{p}%
_{2\pi})}{(2\pi)^6\,\vec{k}_1^2\,\vec{k}_2^2\,\vec{k}_3^2} J_O^{\gamma^*
\rightarrow \pi^+\pi^-}\cdot J_O^{N \rightarrow N^{\prime}}
\end{equation}
where $J_O^{\gamma^* \rightarrow \pi^+\pi^-}(\vec{k}_1,\vec{k}_2,\vec{k}_3)$
and $J_O^{N \rightarrow N^{\prime}}(\vec{k}_1,\vec{k}_2,\vec{k}_2)$ are the
impact factors  for the transition $\gamma^* \to \pi^+\ \pi^-$ via Odderon
exchange  and of the nucleon in initial state $N$ into the nucleon in the
final state $N^{\prime}$.

The upper impact factors are calculated by the use of standard methods, see e.g.
Ref.~\cite{GI} and references therein.

\subsection{Impact factors for $\gamma _{L/T}^{*}\rightarrow \pi ^{+}\pi ^{-}
$}

The leading order calculation in pQCD of the upper impact factors gives in
the case of a
longitudinal polarized photon
\begin{equation}
J_{P}^{\gamma _{L}^{*}}(\vec{k}_{1},\vec{k}_{2})=-\frac{i\,e\,g^{2}\,\delta
^{ab}\,Q}{2\,N_{C}}\;\int\limits_{0}^{1}\,dz\,z{\bar{z}}\,P_{P}(\vec{k}_{1},%
\vec{k}_{2})\,\Phi ^{I=1}(z,\zeta ,m_{2\pi }^{2})  \label{IFPL}
\end{equation}
where $\vec{k}_{1}+\vec{k}_{2}=\vec{p}_{2\pi }$ and the function $P_{P}(\vec{%
k}_{1},\vec{k}_{2})$ is given by
\begin{equation}
P_{P}(\vec{k}_{1},\vec{k}_{2})=\frac{1}{z^{2}\vec{p}_{2\pi }^{\;2}+\mu ^{2}}+%
\frac{1}{{\bar{z}}^{2}\vec{p}_{2\pi }^{\;2}+\mu ^{2}}-\frac{1}{(\vec{k}_{1}-z%
\vec{p}_{2\pi })^{2}+\mu ^{2}}-\frac{1}{(\vec{k}_{1}-{\bar{z}}\vec{p}_{2\pi
})^{2}+\mu ^{2}}
\end{equation}
with $\mu ^{2}=m_{q}^{2}+z\,{\bar{z}}\,Q^{2}$, where $m_{q}$ is the quark
mass and we put $m_{u}\simeq m_{d}=0.006$~GeV. The GDA $\Phi
^{I=0,1}(z,\zeta ,m_{2\pi }^{2})$ for isospin $I=0,1$ will be discussed in
detail in the next section. The computation of the three-gluon-exchange
graphs for the longitudinally polarized photon results in the following impact
factor
\begin{equation}
J_{O}^{\gamma _{L}^{*}}(\vec{k}_{1},\vec{k}_{2},\vec{k}_{3})=-\frac{%
i\,e\,g^{3}\,d^{abc}\,Q}{4\,N_{C}}\;\int\limits_{0}^{1}\,dz\,z{\bar{z}}%
\,P_{O}(\vec{k}_{1},\vec{k}_{2},\vec{k}_{3})\,\frac{1}{3}\Phi ^{I=0}(z,\zeta
,m_{2\pi }^{2})  \label{IFOL}
\end{equation}
where $\vec{k}_{1}+\vec{k}_{2}+\vec{k}_{3}=\vec{p}_{2\pi}$ and
\begin{eqnarray}
&&P_{O}(\vec{k}_{1},\vec{k}_{2},\vec{k}_{3})=\frac{1}{z^{2}\vec{p}_{2\pi
}^{\;2}+\mu ^{2}}-\frac{1}{{\bar{z}}^{2}\vec{p}_{2\pi }^{\;2}+\mu ^{2}}
\nonumber \\
&&-\sum\limits_{i=1}^{3}\left( \frac{1}{(\vec{k}_{i}-z\vec{p}_{2\pi
})^{2}+\mu ^{2}}-\frac{1}{(\vec{k}_{i}-{\bar{z}}\vec{p}_{2\pi })^{2}+\mu ^{2}%
}\right)
\end{eqnarray}
In the case of a transversely polarized photon we introduce the transverse
photon polarization vectors $\vec{\epsilon}(T=+,-)$ through
\begin{equation}
\vec{\epsilon}(+)=-\frac{1}{\sqrt{2}}(1,i),\,\;\;\;\;\vec{\epsilon}(-)=\frac{1%
}{\sqrt{2}}(1,-i).
\end{equation}
Using this, the upper impact factor for the Pomeron induced process can be
written as
\begin{equation}
J_{P}^{\gamma _{T}^{*}}(\vec{k}_{1},\vec{k}_{2})=-\frac{i\,e\,g^{2}\,\delta
^{ab}}{4\,N_{C}}\,\int\limits_{0}^{1}\,dz\,(z-{\bar{z}})\;\vec{\epsilon}%
(T)\cdot \vec{Q}_{P}(\vec{k}_{1},\vec{k}_{2})\;\Phi ^{I=1}(z,\zeta ,m_{2\pi
}^{2})  \label{IFPT}
\end{equation}
where the vector $\vec{Q}_{P}(\vec{k}_{1},\vec{k}_{2})$ is defined by
\begin{eqnarray}
\label{QP}
\vec{Q}_{P}(\vec{k}_{1},\vec{k}_{2})= 
=\frac{z\vec{p}_{2\pi }}{z^{2}\vec{p}_{2\pi }^{\;2}+\mu ^{2}}-\frac{{\bar{z}}%
\vec{p}_{2\pi }}{{\bar{z}}^{2}\vec{p}_{2\pi }^{\;2}+\mu ^{2}}+\frac{\vec{k}%
_{1}-z\vec{p}_{2\pi }}{(\vec{k}_{1}-z\vec{p}_{2\pi })^{2}+\mu ^{2}}-\frac{%
\vec{k}_{1}-{\bar{z}}\vec{p}_{2\pi }}{(k_{1}-{\bar{z}}\vec{p}_{2\pi
})^{2}+\mu ^{2}}  
\end{eqnarray}
The calculation of the Odderon exchange contribution gives 
\begin{equation}
J_{O}^{\gamma _{T}^{*}}(\vec{k}_{1},\vec{k}_{2},\vec{k}_{3})=-\frac{%
i\,e\,g^{3}\,d^{abc}}{8\,N_{C}}\,\int\limits_{0}^{1}\,dz\,(z-{\bar{z}})\;%
\vec{\epsilon}(T)\cdot \vec{Q}_{O}(\vec{k}_{1},\vec{k}_{2},\vec{k}_{3})\;%
\frac{1}{3}\Phi ^{I=0}(z,\zeta ,m_{2\pi }^{2})  \label{IFOT}
\end{equation}
where we have used the definition
\begin{eqnarray}
\label{QO}
&&\vec{Q}_{O}(\vec{k}_{1},\vec{k}_{2},\vec{k}_{3})= \\
&=&\frac{z\vec{p}_{2\pi }}{z^{2}\vec{p}_{2\pi }^{\;2}+\mu ^{2}}+\frac{{\bar{z}}%
\vec{p}_{2\pi }}{{\bar{z}}^{\;2}\vec{p}_{2\pi }^{2}+\mu ^{2}}%
+\sum\limits_{i=1}^{3}\left( \frac{\vec{k}_{i}-z\vec{p}_{2\pi }}{(\vec{k}%
_{i}-z\vec{p}_{2\pi })^{2}+\mu ^{2}}+\frac{\vec{k}_{i}-{\bar{z}}\vec{p}%
_{2\pi }}{(k_{i}-{\bar{z}}\vec{p}_{2\pi })^{2}+\mu ^{2}}\right)   \nonumber
\end{eqnarray}
The value of the strong coupling constant $g$ in the hard block is assumed
to correspond to the 1-loop running coupling constant with $n_{f}=2$, $%
\alpha _{s}(Q^{2})=\frac{g^{2}}{4\pi }={12\pi }/[{29\ln (\frac{Q^{2}}{%
\Lambda _{QCD}^{2}})}]$. In our numerical estimates we take as a mean value $%
\Lambda _{QCD}=0.25$ GeV.
Varying this value in a reasonable range does not modify much our results.

\subsection{Generalized two pion distribution amplitudes}

A crucial point of the present study is the choice of an appropriate
two pion distribution amplitude (GDA) \cite{DGPT,POL,DGP} which includes
the full strong interaction related to the production of the two
pion system.
We follow here the discussion in our previous paper~\cite{HPST} and 
propose a possible improvement of this GDA in section 5.

The Odderon induced contribution we are looking for  is directly proportional to
the $I=0$ part of the GDA, for which we use the following approximation
\begin{eqnarray}
&&\Phi ^{I=0}(z,\zeta ,m_{2\pi })=10z\bar{z}(z-\bar{z})\,R_{\pi }\,
\nonumber \\
&&\left[ -\frac{3-\beta ^{2}}{2}\,e^{i\delta _{0}(m_{2\pi })}\
|BW_{f_{0}}(m_{2\pi }^{2})|+\beta ^{2}\ e^{i\delta _{2}(m_{2\pi })}\
|BW_{f_{2}}(m_{2\pi }^{2})|\ P_{2}(\cos \theta )\right] ,  \label{GDAI0}
\end{eqnarray}
with $R_{\pi }=0.5$ and $\beta $ given by Eq. (\ref{theta}). 
In our studies we fix the shapes of the phase shifts $\delta _{0}$ and $%
\delta _{2}$ by a fit to data presented in \cite{Hyams}. The factors $%
|BW_{f_{0,2}}(m_{2\pi }^{2})|$ are the modulus of the Breit-Wigner
amplitudes
\begin{equation}
BW_{f_{0}}(m_{2\pi }^{2})=\frac{m_{f_{0}}^{2}}{m_{f_{0}}^{2}-m_{2\pi
}^{2}-im_{f_{0}}\Gamma _{f_{0}}}\ ,\;\;\;\;m_{f_{0}}=0.98\;\text{GeV}%
,\;\;\;\;\Gamma _{f_{0}}=0.075\;\text{GeV}  \label{f0}
\end{equation}
\begin{equation}
BW_{f_{2}}(m_{2\pi }^{2})=\frac{m_{f_{2}}^{2}}{m_{f_{2}}^{2}-m_{2\pi
}^{2}-im_{f_{2}}\Gamma _{f_{2}}}\ ,\;\;\;\;m_{f_{2}}=1.275\;\text{GeV}%
\,,\;\;\;\;\Gamma _{f_{2}}=0.186\;\text{GeV}\ .  \label{f2}
\end{equation}
Modifying the $f_0$ width changes slightly our results, as discussed in Ref. \cite{HPST}.

For the isospin $I=1$ part of the two pion GDA, which is relevant for the
Pomeron exchange amplitude, we take
\begin{equation}
\Phi ^{I=1}(z,\zeta ,m_{2\pi })=6z\bar{z}\beta \cos \theta \;F_{\pi
}(m_{2\pi }^{2})\ ,  \label{GDAI1}
\end{equation}
where the timelike pion form factor is parametrized by
\begin{equation}
F_{\pi }(m_{2\pi }^{2})=\frac{1}{(1-0.145)}BW_{\rho }\,\frac{1+1.85\cdot
10^{-3}\cdot BW_{\omega }}{1+1.85\cdot 10^{-3}}\ ,  \label{FFrho'}
\end{equation}
with
\begin{eqnarray}
&&BW_{\rho }(m_{2\pi }^{2})=\frac{m_{\rho }^{2}}{m_{\rho }^{2}-m_{2\pi
}^{2}-i\sqrt{m_{2\pi }^{2}}\Gamma _{\rho }(m_{2\pi }^{2})}\ , \\
&&\Gamma _{\rho }(m_{2\pi }^{2})=\Gamma _{\rho }\,\frac{m_{\rho }^{2}}{%
m_{2\pi }^{2}}\,\frac{(m_{2\pi }^{2}-4\,m_{\pi }^{2})^{3/2}}{(m_{\rho
}^{2}-4\,m_{\pi }^{2})^{3/2}}\ ,\;\;\;m_{\rho }=0.773\;\text{GeV},\;\;\Gamma
_{\rho }=0.145\;\text{GeV}\,  \nonumber
\end{eqnarray}
and
\begin{equation}
BW_{\omega }(m_{2\pi }^{2})=\frac{m_{\omega }^{2}}{m_{\omega }^{2}-m_{2\pi
}^{2}-im_{\omega }\Gamma _{\omega }}\ ,\;\;\;\;m_{\omega }=0.782\;\text{GeV}%
,\;\;\Gamma _{\omega }=0.0085\;\text{GeV}.
\end{equation}
The phase of the form factor $F_{\pi}(m_{2\pi }^2)$ will be denoted by $e^{i\delta_1}$,
where $\delta_1$ is the corresponding $p-$wave phase shift.

\subsection{Proton impact factors}

Finally we have to fix the lower soft parts of our amplitudes, i.e. the proton
impact factors. They cannot be calculated within perturbation theory. In our
estimates we will use phenomenological eikonal models of these impact
factors proposed in Refs. \cite{protonP} and \cite{protonO}. We take
for the Pomeron exchange

\begin{equation}
J_{P}^{N\rightarrow N^{\prime }}=i\frac{{\bar{g}}^{2}\,\delta ^{ab}}{2\,N_{C}%
}\,3\,\left[ \frac{A^{2}}{A^{2}+\frac{1}{2}\,\vec{p}_{2\pi }^{2}}-\frac{A^{2}%
}{A^{2}+\frac{1}{2}(\vec{k}_{1}^{2}+\vec{k}_{2}^{2})}\right] \,,\;\;
\label{IFPP}
\end{equation}
and for the Odderon exchange
\begin{equation}
\label{IFPO}
J_{O}^{N\rightarrow N^{\prime }}=-i\frac{{\bar{g}}^{3}\,d^{abc}}{4\,N_{C}}%
\,3\left[ F(\vec{p}_{2\pi },0,0)-\sum\limits_{i=1}^{3}F(\vec{k}_{i},\vec{p}%
_{2\pi }-\vec{k}_{i},0)+2\,F(\vec{k}_{1},\vec{k}_{2},\vec{k}_{3})\right]
\end{equation}
where
\begin{equation}
F(\vec{k}_{1},\vec{k}_{2},\vec{k}_{3})=\frac{A^{2}}{A^{2}+\frac{1}{2}\left[ (%
\vec{k}_{1}-\vec{k}_{2})^{2}+(\vec{k}_{2}-\vec{k}_{3})^{2}+(\vec{k}_{3}-\vec{%
k}_{1})^{2}\right] }
\end{equation}
and $A=\frac{m_{\rho }}{2}$.
In these equations we have denoted the soft QCD-coupling constant by ${\bar
g}$.
We take ${\alpha}_{soft} =\bar{g}^2/(4\pi)=0.5$ as a reasonable mean value (for discussion
of this point see our paper \cite{HPST}).



\begin{figure}[t]
\begin{minipage}[t]{75mm}
\centerline{\includegraphics[scale=.38]{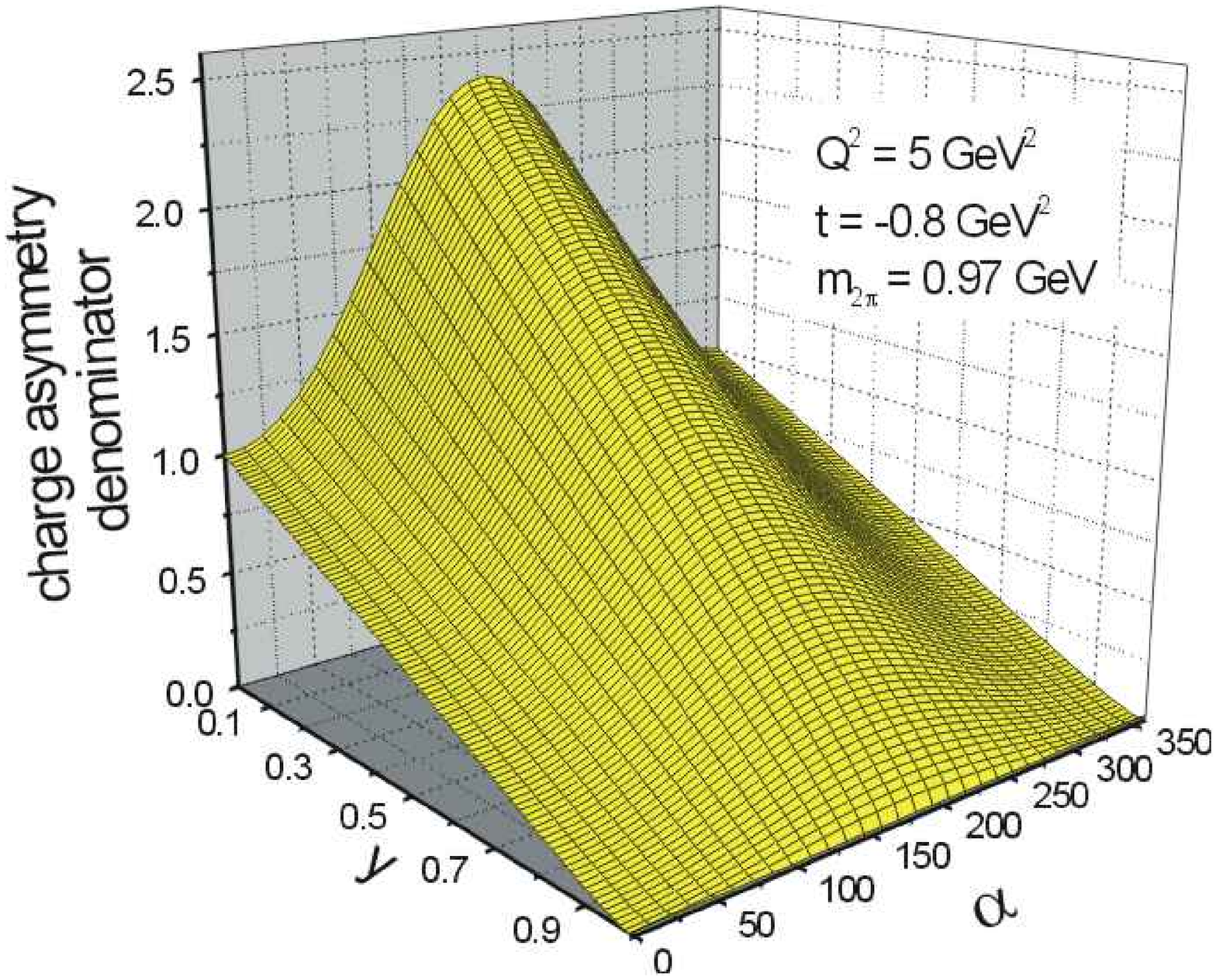}}
\caption{{\protect\small $(y,\alpha)$ dependence of the denominator
(\ref{denca}) of the asymmetries for
$m_{2\pi}=0.97$ GeV, $t=-0.8$ GeV$^2$ and $Q^2=5$ GeV$^2$
}}
\label{denom}
\end{minipage}
\hspace{\fill}
\begin{minipage}[t]{75mm}
\centerline{\includegraphics[scale=.6]{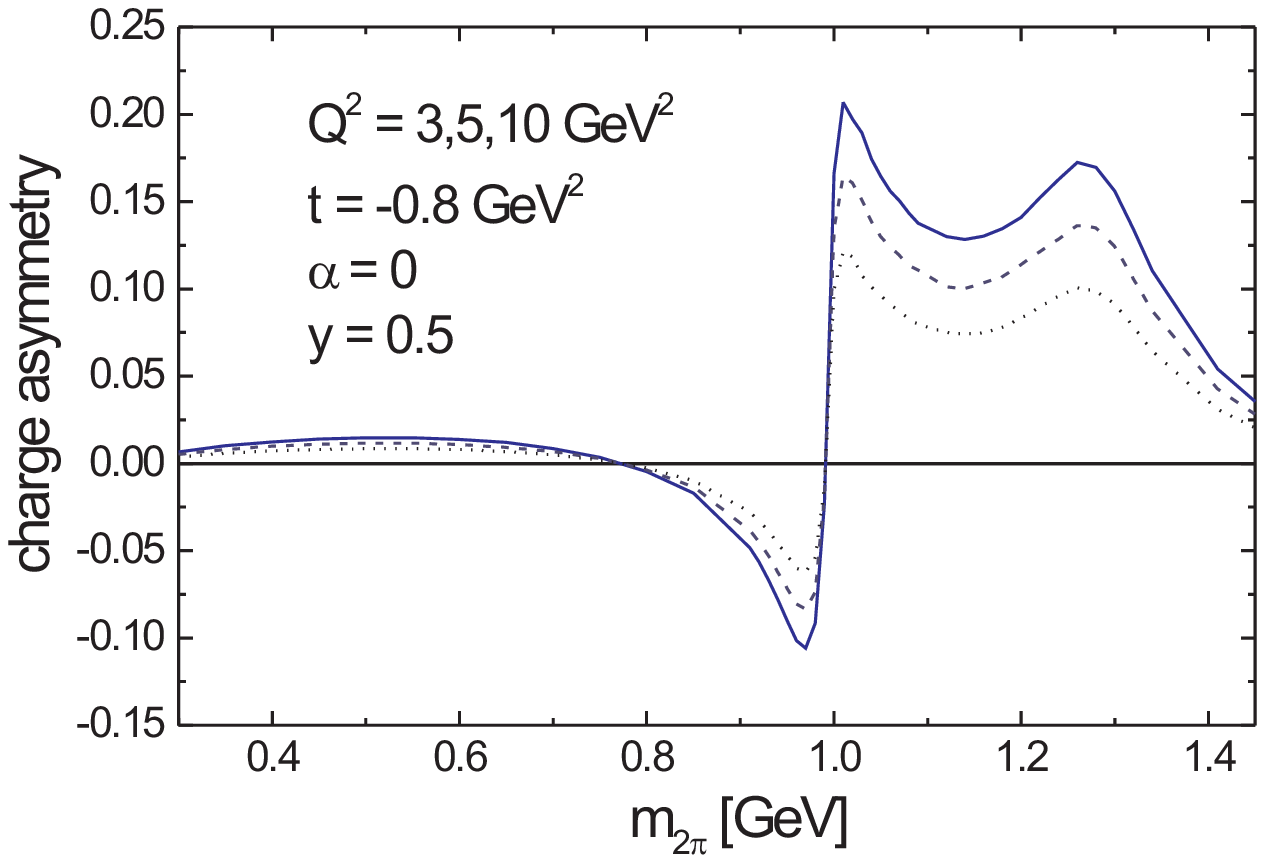}}
\caption{{\protect\small $m_{2\pi}$ dependence of the charge asymmetry for
$Q^2=$
3 GeV$^2$ (solid line), 5 GeV$^2$ (dashed line) and 10 GeV$^2$ (dotted line)
for $t=-0.8$GeV$^2$, $\alpha=0$ and $y=0.5$ }}
\label{m2pi}
\end{minipage}
\end{figure}


\section{Asymmetries and their numerical evaluation}

Taking together equations (\ref{pom}-\ref{IFPL}, \ref{IFOL}, \ref{IFPT},
\ref{IFOT}, 
\ref{GDAI0}, \ref{GDAI1}, \ref{IFPP}, \ref{IFPO}) we are now ready for the
calculation of the asymmetries and their subsequent numerical evaluation. 
In contrast to the results presented in \cite{HPST} where we only considered
the charge asymmetry resulting from the scattering of a longitudinally polarized photon,  
we consider below the contributions to the charge asymmetry comming from both longitudinal
and transverse photon degrees of freedom.
Moreover, we study the single spin asymmetry which involves amplitudes 
with transversely and longitudinally polarized photons.

Because all photon polarizations contribute to the asymmetries they will
 now depend on the
angle $\alpha $ between the initial electron transverse momentum $\p_i$ and the 
tranverse momentum of the pion pair $\p_{2\pi}$,
 and on
the energy loss $y$ (Eq.~(\ref{y})) of the initial electron.

\subsection{Charge asymmetry}




We define the  forward-backward or charge asymmetry by

\begin{equation}
\label{ca}
A(Q^{2},t,m_{2\pi }^{2},y,\alpha )=\frac{\sum\limits_{\lambda =+,-}\int \cos
\theta \,d\sigma (s,Q^{2},t,m_{2\pi }^{2},y,\alpha ,\theta ,\lambda )}{%
\sum\limits_{\lambda =+,-}\int d\sigma (s,Q^{2},t,m_{2\pi }^{2},y,\alpha
,\theta ,\lambda )}=\frac{\int d\cos
\theta \cos\theta \;N_{charge}}{\int d\cos
\theta \;D}  
\end{equation}
We observe that the vectors $\vec{Q}_{P/O}$ in Eqs.~(\ref{QP}, \ref{QO}), after
integration over the
gluon momenta $\vec{k}_{i}$, can be only proportional to $\vec{p}_{2\pi }$.
Therefore it is useful to define scalar functions ${\cal A}_{T}(P/O)$ by
\begin{equation}
\nonumber
 \vec{\epsilon}%
(T)\cdot \vec{p}_{2\pi }\,{\cal A}_{T}(P/O)\equiv \vec{\epsilon}(T)\cdot
\vec{Q}%
_{P/O}.
\end{equation}
Using this, the calculation of the numerator $N_{charge}$ and the denominator
$D$ gives
\begin{eqnarray}
\label{nomca}
N_{charge} &=&8(1-y)\func{Re}\left[ \mathcal{M}_{L}(P)\mathcal{M}%
_{L}^{*}(O)\right]  \\
&&+4(2-y)\sqrt{1-y}|\vec{p}_{2\pi }|\cos \alpha \;\func{Re}\left[
{\cal A}_{T}(P)%
\mathcal{M}_{L}^{*}(O)+{\cal A}_{T}(O)\;\;\mathcal{M}_{L}^{*}(P)\right]
\nonumber
\\
&&+2(1+(1-y)^{2}+2(1-y)\cos 2\alpha ) |\vec{p}_{2\pi }|^{2}\func{Re}%
\left[ {\cal A}_{T}(P){\cal A}_{T}^{*}(O)\right]   \nonumber
\end{eqnarray}
and
\begin{eqnarray}
\label{denca}
D &=&4(1-y)\left| \mathcal{M}_{L}(P)+\mathcal{M}_{L}(O)\right| ^{2} \\
&&+4(2-y)\sqrt{1-y}|\vec{p}_{2\pi }|\cos \alpha \;\func{Re}\left[ \left(
{\cal A}_{T}(P)+{\cal A}_{T}(O)\right) \left(
\mathcal{M}_{L}^{*}(P)+\mathcal{M}%
_{L}^{*}(O)\right) \right]   \nonumber \\
&&+(1+(1-y)^{2}+2(1-y)\cos 2\alpha)|\vec{p}_{2\pi }|^{2}\left|
{\cal A}_{T}(P)+{\cal A}_{T}(O)\right| ^{2}  \nonumber
\end{eqnarray}


\begin{figure}[t]
\begin{minipage}[t]{75mm}
\centerline{\includegraphics[scale=.38]{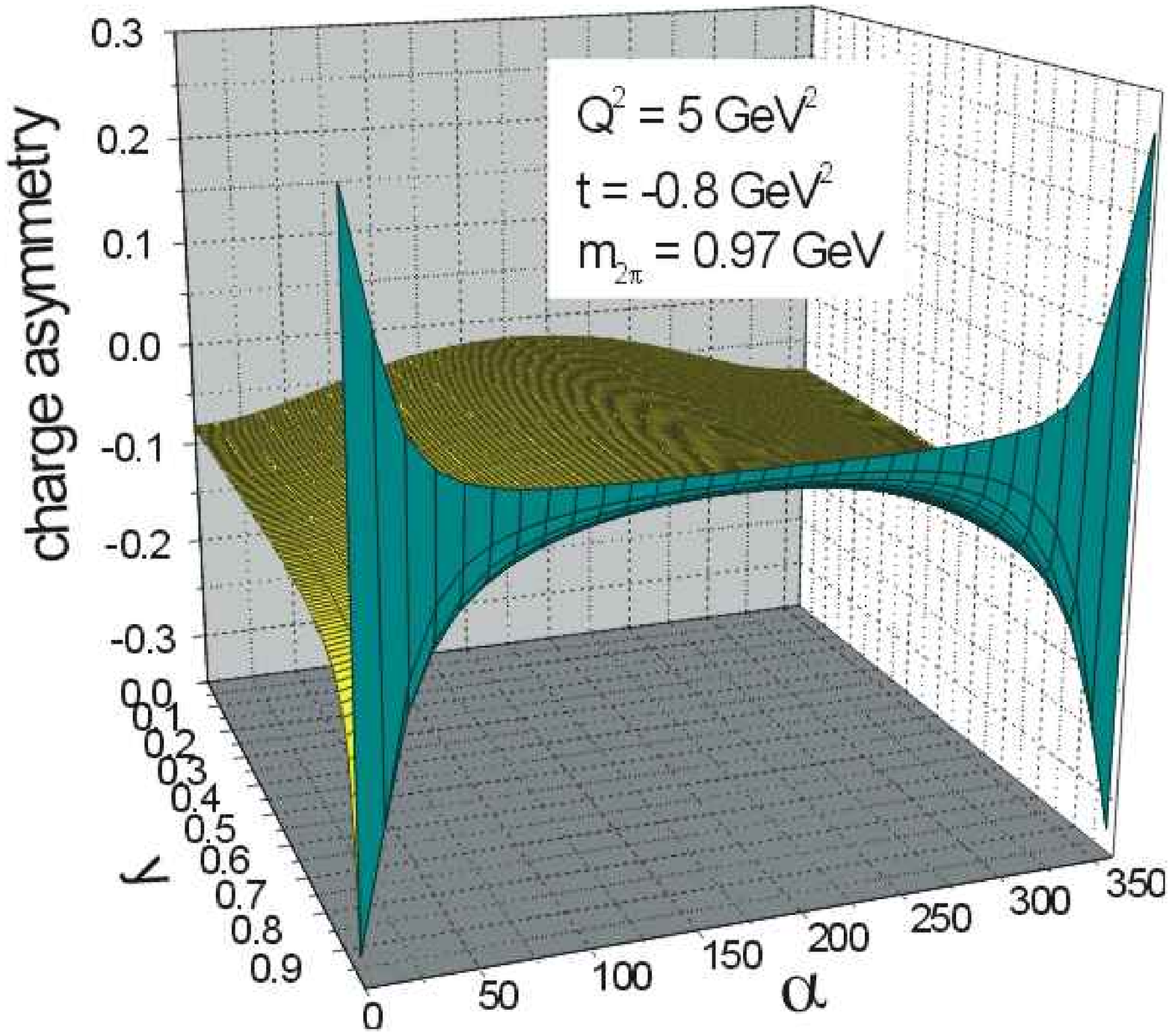}}
\caption{{\protect\small $(y,\alpha)$ dependence of the charge asymmetry for
$m_{2\pi}=0.97$ GeV, $t=-0.8$ GeV$^2$ and $Q^2=5$ GeV$^2$ seen from the
$\alpha$  side }}
\label{3da2}
\end{minipage}
\hspace{\fill}
\begin{minipage}[t]{75mm}
\centerline{\includegraphics[scale=.38]{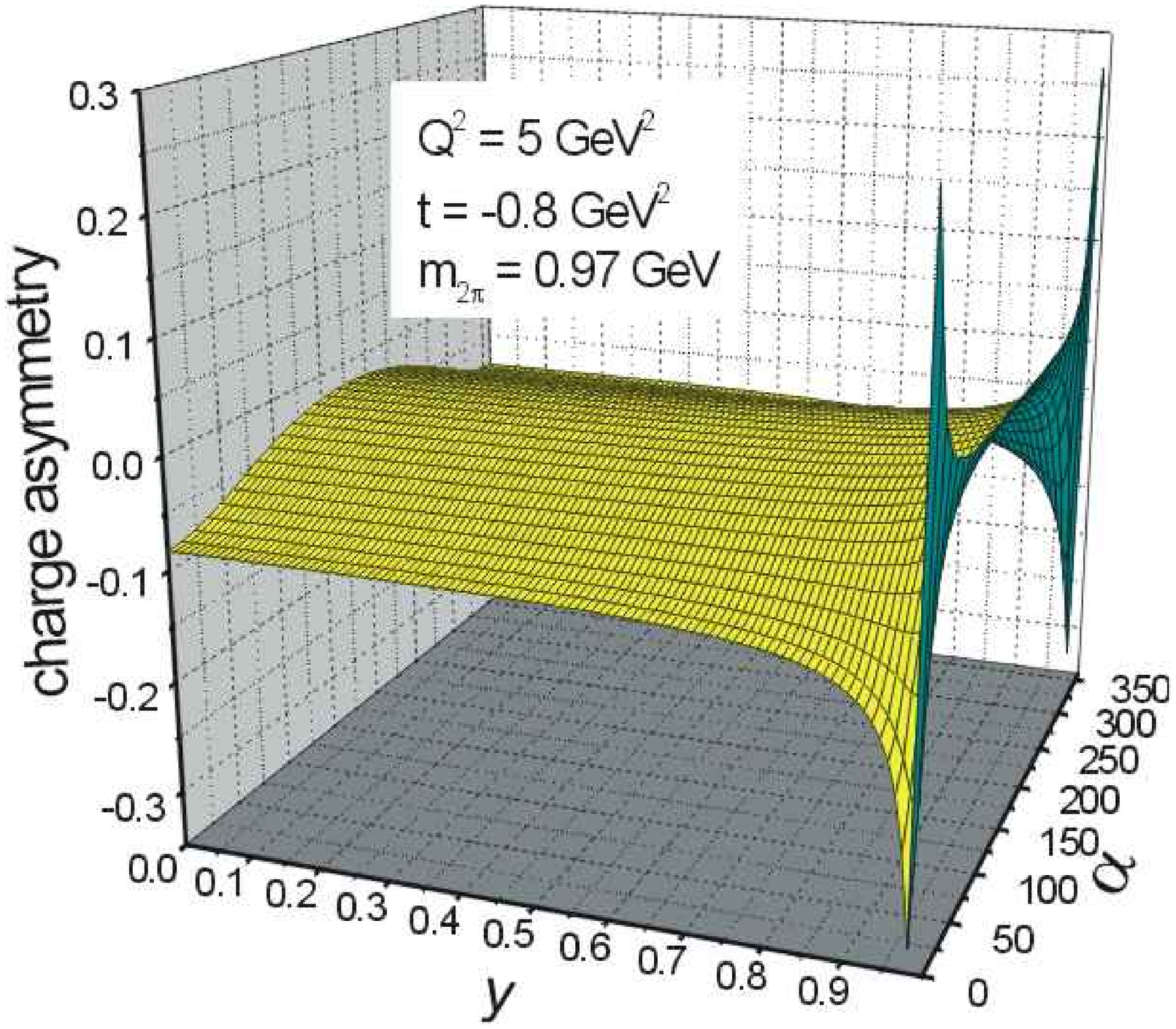}}
\caption{{\protect\small $(y,\alpha)$ dependence of the charge asymmetry for
$m_{2\pi}=0.97$ GeV, $t=-0.8$ GeV$^2$ and $Q^2=5$ GeV$^2$ seen from the
$y$  side }}
\label{3dy2}
\end{minipage}
\end{figure}

%

\begin{figure}[t]
\begin{minipage}[t]{75mm}
\centerline{\includegraphics[scale=.38]{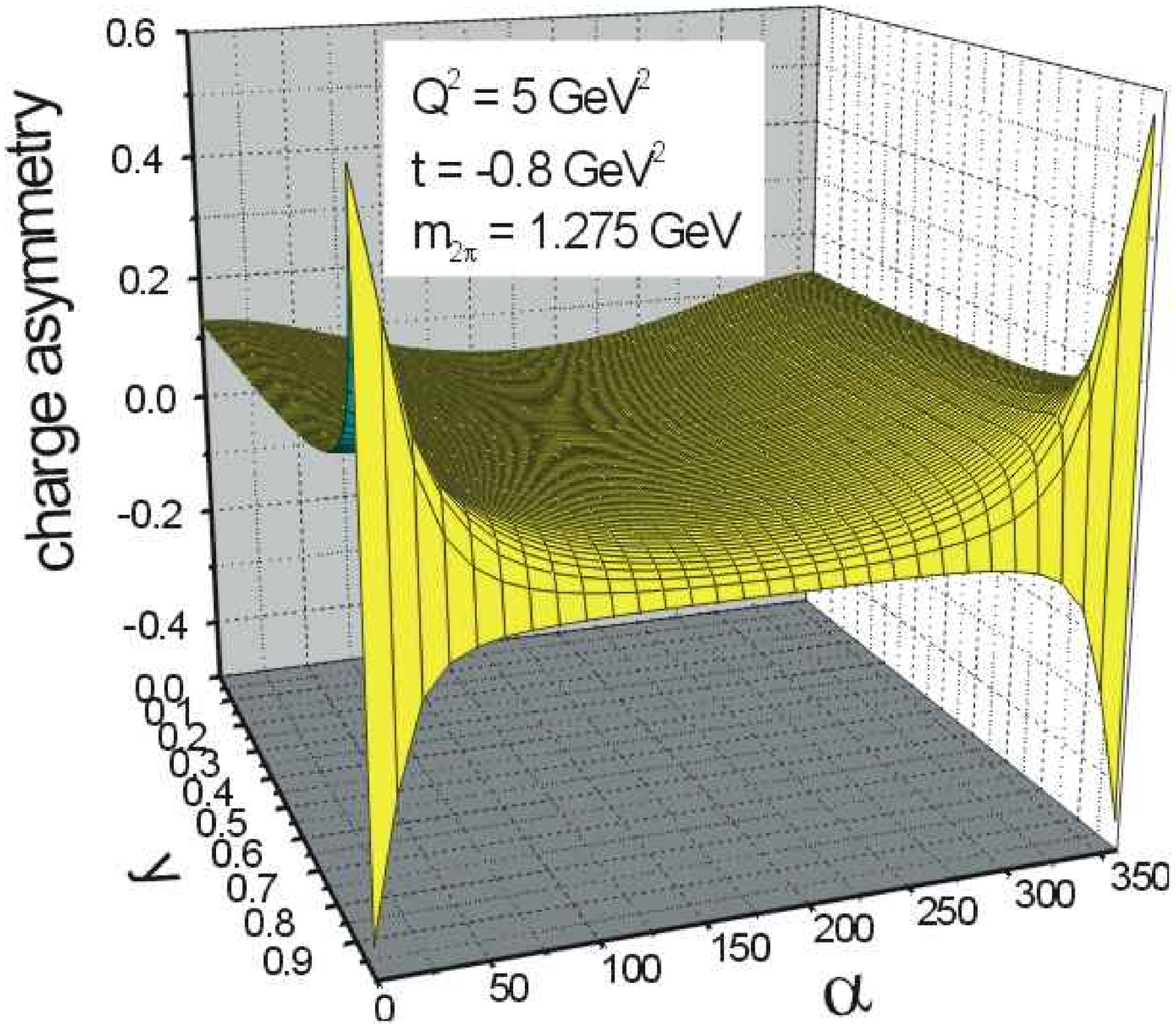}}
\caption{{\protect\small $(y,\alpha)$ dependence of the charge asymmetry for
$m_{2\pi}=1.275$ GeV, $t=-0.8$ GeV$^2$ and $Q^2=5$ GeV$^2$ seen from the
$\alpha$  side }}
\label{3da}
\end{minipage}
\hspace{\fill}
\begin{minipage}[t]{75mm}
\centerline{\includegraphics[scale=.38]{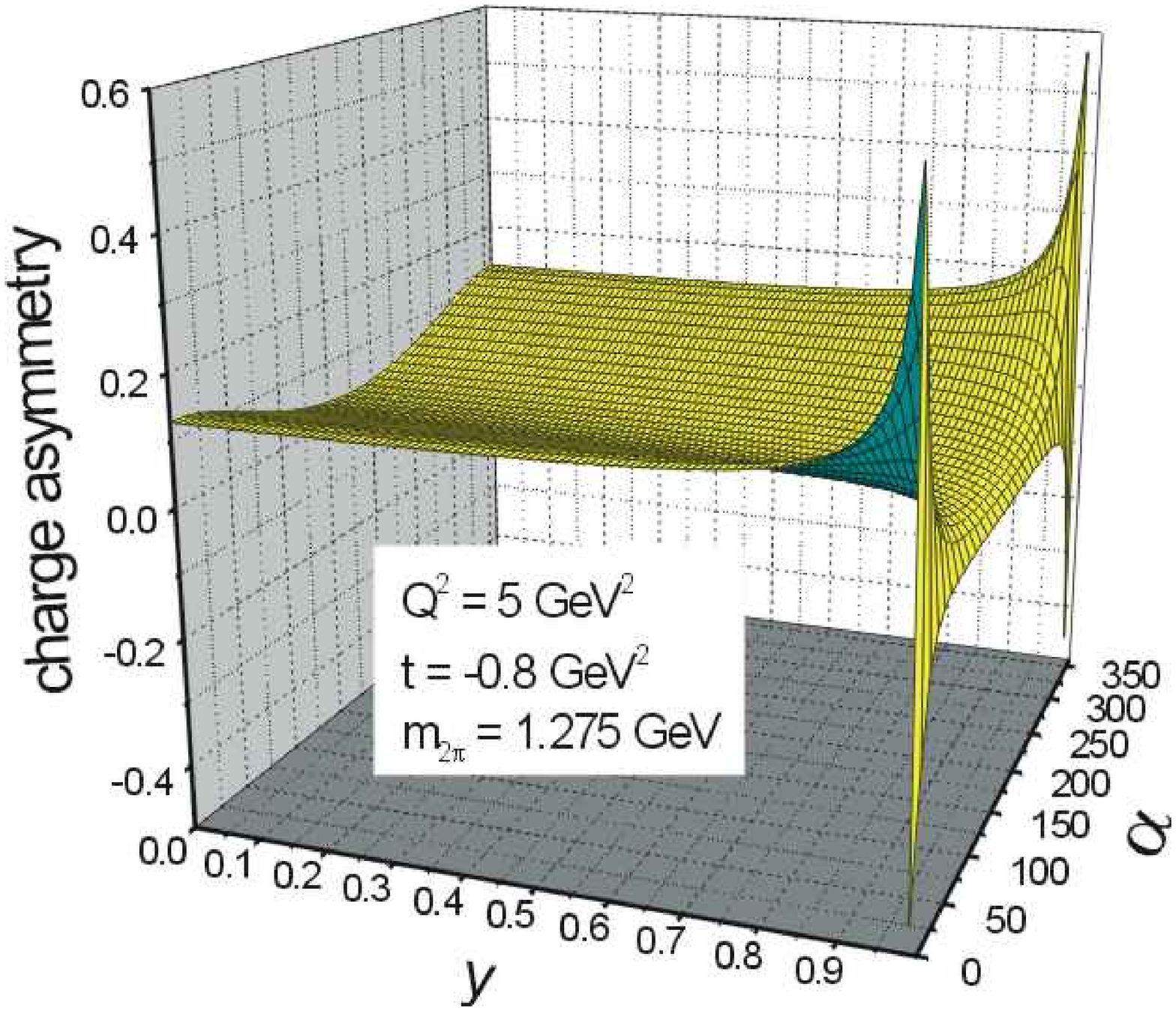}}
\caption{{\protect\small $(y,\alpha)$ dependence of the charge asymmetry for
$m_{2\pi}=1.275$ GeV, $t=-0.8$ GeV$^2$ and $Q^2=5$ GeV$^2$ seen from the
$y$  side }}
\label{3dy}
\end{minipage}
\end{figure}





Instead of a weighted integration of the cross-section over $\theta$ it is
possible to perform a full angular analysis. The numerator of the asymmetry
would then be  provided by the $\cos \theta$-term  which is characteristic
for the longitudinal polarization of the pion pair.

We checked that the squared Odderon contribution in the denominator can be
neglected (except for the large $y$ region),  so that the
asymmetry is practically a measure of the ratio of the Odderon and the
Pomeron amplitudes.


%

Before presenting our results for the asymmetries and in order to get some
handle on relative counting rates we show in Fig. \ref{denom}
the $\alpha$ and $y$ behaviour of the denominator $D$ (in arbitrary
units), which is directly proportional to the charge averaged differential cross section.
One sees clearly that the denominator is only about a factor of 2-4
below its maximum value at $y=0$ and $\alpha =\pi/2$ in a region which
is experimentally accessible and where the asymmetry to be discussed
below is relatively large.


The characterisic $m_{2\pi }$ dependence shown in Fig. \ref{m2pi} is
completely understood in terms of the $\pi \pi $ phase shifts and the factor
$\sin (\delta_{0,2}-\delta_{1})$. The phase difference vanishes for $%
m_{2\pi }\approx 0.75\ $GeV and $m_{2\pi }\approx 1\ $GeV resulting in two
zeros of the charge asymmetry. The magnitude of the charge asymmetry is
quite large around the $f_{0}$ and $f_{2}$ masses. It depends somewhat on
the width of the $f_{0}$ meson which is taken to be $0.075$ GeV (see the
discussion in Ref. \cite{HPST}). The $Q^{2}$ dependence of the charge
asymmetry is, as seen from Fig.~\ref{m2pi}, moderate but of course the cross
section increases with decreasing $Q^{2}$.

The $\alpha $ and $y$ dependences of the charge asymmetry are shown in Figs.   
\ref{3da2} and \ref{3dy2} for a value of $m_{2\pi }$ just below the
$f_{0}$ mass, and on Figs. \ref{3da}, \ref{3dy}
 for a value of $m_{2\pi }$ just equal to the $f_{2}$ mass,
 where the
asymmetry is large. The effect is maximal for values of 
$\alpha \approx 0$ and
minimal for $\alpha \sim \pi $. The dependence on $y$ is very weak except
for the region  $y\rightarrow 1$ where the cross section is so small that
no experimental data will ever be  available.

\begin{figure}[t]
\begin{minipage}[t]{75mm}
\centerline{\includegraphics[scale=.6]{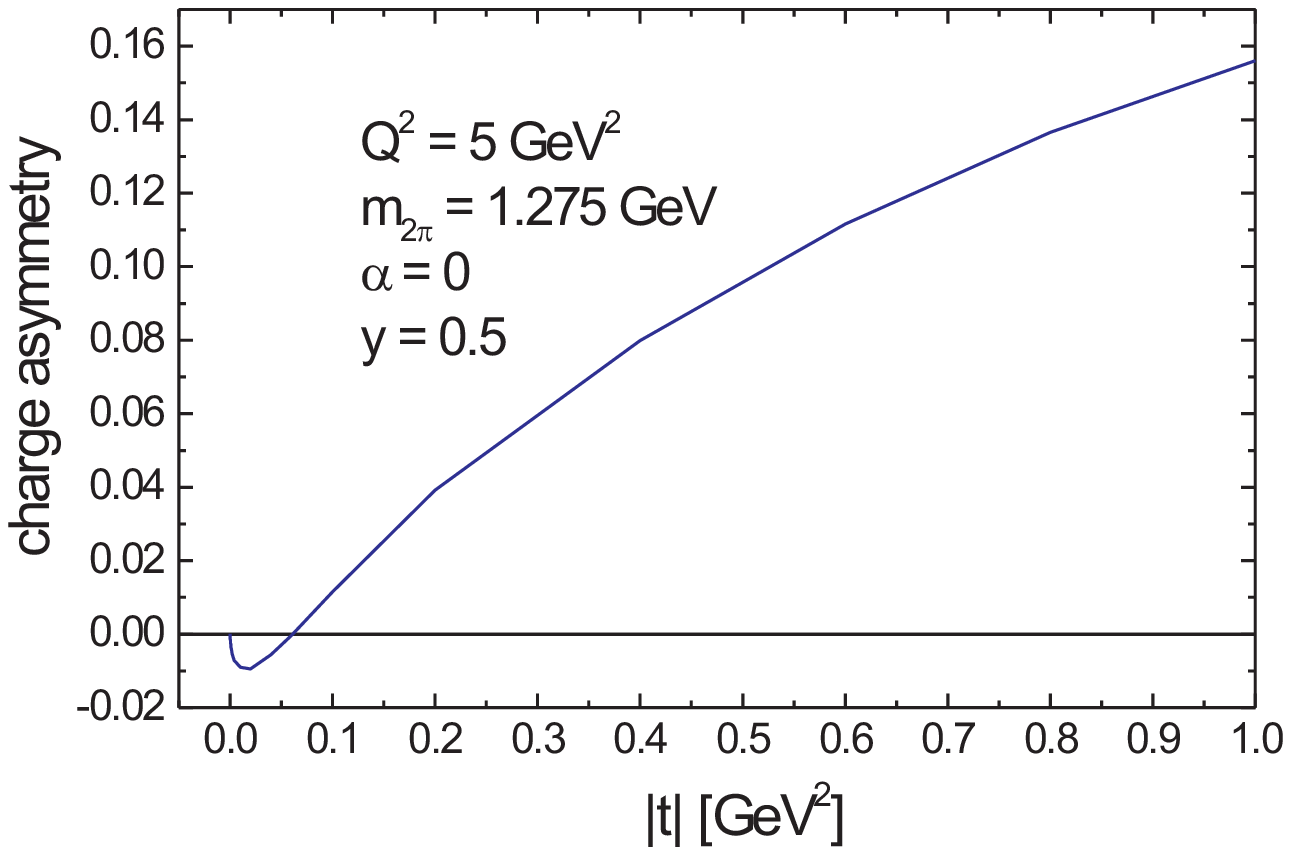}}
\caption{{\protect\small $t-$dependence of the charge asymmetry at $m_{2\pi}=
1.275$ GeV, $Q^2=5$ GeV$^2$ }}
\label{t}
\end{minipage}
\hspace{\fill}
\begin{minipage}[t]{75mm}
\centerline{\includegraphics[scale=0.6]{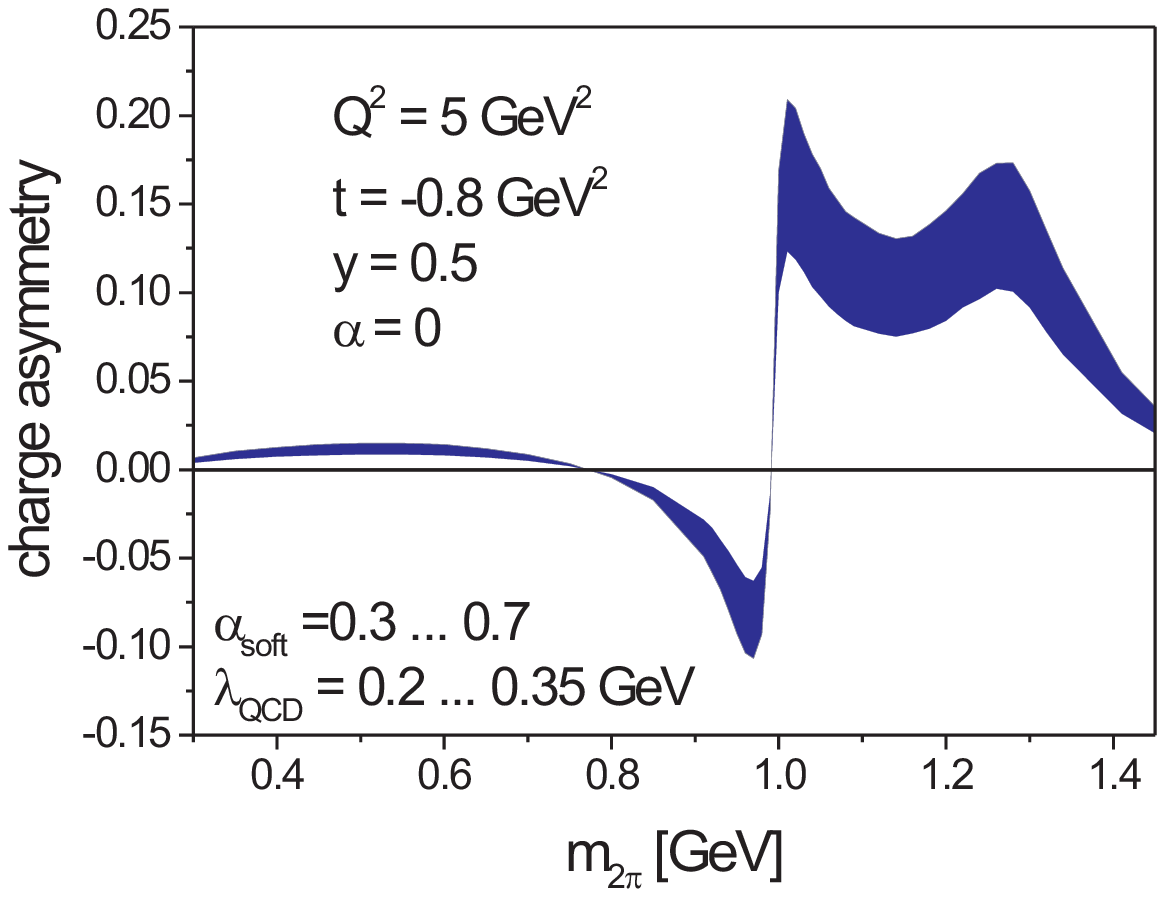}}
\caption{{\protect\small Error bands resulting from a variation of
$\Lambda_{QCD}$
and the soft coupling $\alpha_{soft}$.}}
\label{errorband}
\end{minipage}
\end{figure}

The $t$ dependence of the asymmetry is plotted in Fig.~\ref{t}. It has a
characteristic zero around $t=-0.06$ GeV$^{2}$. This zero in the odderon
amplitude has already been discussed in Ref. \cite{Vacca2}.

In Fig.\ref{errorband} we show an error band for the
$m_{2\pi}$-dependent charge asymmetry resulting from a simultaneous
variation of $\Lambda_{QCD}$ and the soft coupling $\alpha_{soft}$
in the indicated range.

\subsection{ Spin asymmetry}

The presence of the interference between the different helicity
amplitudes with non-zero phase shift between them provides
the necessary conditions for the emergence of single spin asymmetries.

%
The most realistic is the single spin asymmetry
generated by the polarized lepton beam, which is
accessible at HERA. The resulting azimuthal
asymmetry is analogous
to the ones measured at lower energies by HERMES
and CLAS collaborations \cite{Azim}.

The single spin asymmetry is defined by

\begin{eqnarray}
\label{sa}
&&A_{S}(Q^{2},t,m_{2\pi }^{2},y,\alpha )=\frac{\sum\limits_{\lambda
=+,-}\lambda \int \cos \theta \,d\sigma (s,Q^{2},t,m_{2\pi }^{2},y,\alpha
,\theta ,\lambda )}{\ \sum\limits_{\lambda =+,-}\int d\sigma
(s,Q^{2},t,m_{2\pi }^{2},y,\alpha ,\theta ,\lambda )}=\frac{\int d\cos
\theta \cos\theta \;N_{spin}}{\int d\cos
\theta \;D} \nonumber \\
&&\mbox{}
\end{eqnarray}
and the calculation of the numerator gives
\begin{equation}
\label{nomsa}
N_{spin}=4y\sqrt{1-y}\sin \alpha \;|\vec{p}_{2\pi }|\;\func{Im}\left[
\mathcal{M}_{L}(P)\mathcal{A}%
_{T}^{*}(O)+\mathcal{M}_{L}(O)\mathcal{A}_{T}^{*}(P)\right]
\end{equation}
while $D$ is of course the same quantity as in the case of the charge asymmetry
(\ref{denca}).
In order to increase the magnitude of the spin asymmetry we defined it, by analogy
to the charge asymmetry (\ref{ca}), with an integration over the angle
$\theta$ (or the variable $\zeta$, see Eq.(\ref{theta})) weighted with $\cos \theta$.
The integration over
$\theta$ without a weight factor gives in our approach zero.

\begin{figure}[t]
\begin{minipage}[t]{75mm}
\centerline{\includegraphics[scale=0.6]{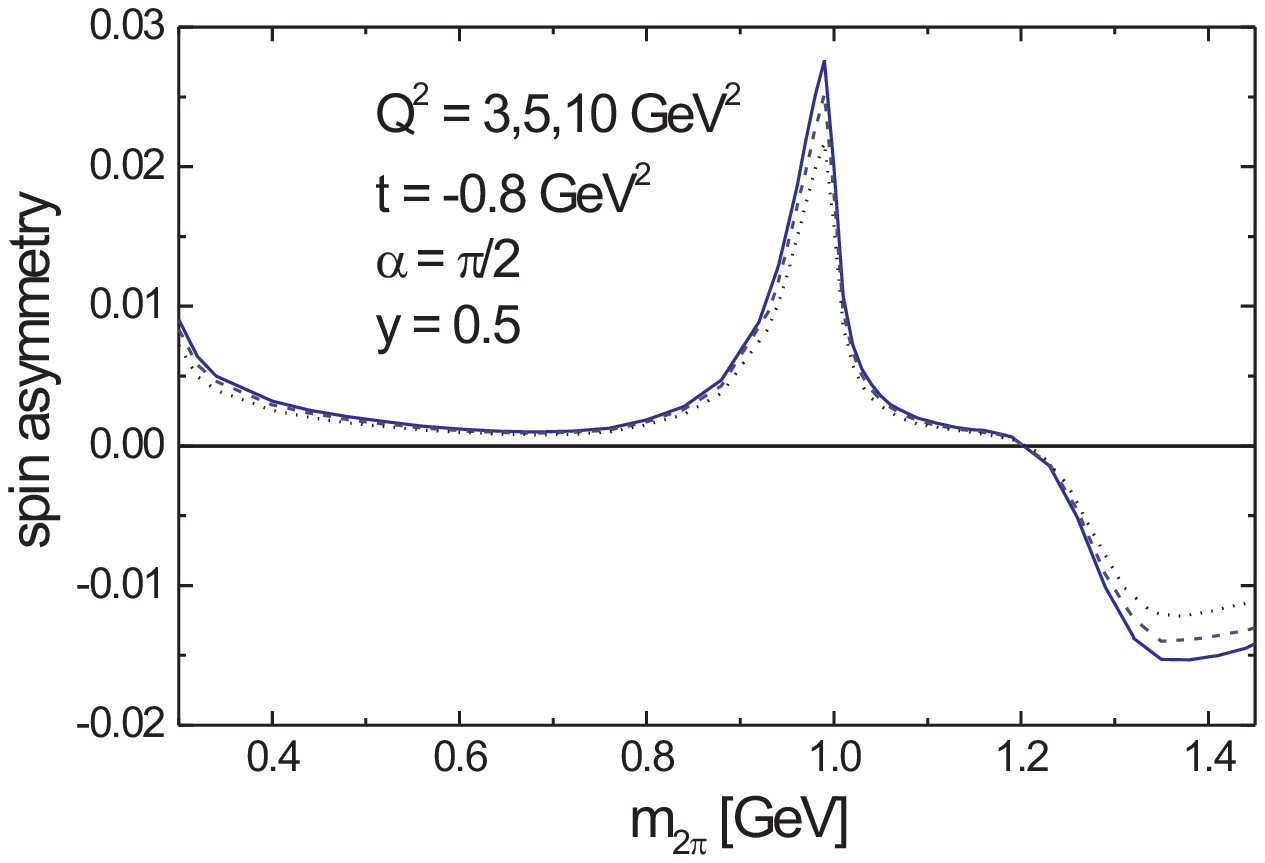}}
\caption{{\protect\small $m_{2\pi}-$dependence of the spin asymmetry at
$t=-0.8$
 GeV$^2$, $Q^2=3$ (solid line), 5 (dashed line), 10 (dotted line)  GeV$^2$ }}
\label{m2pispin}
\end{minipage}
\hspace{\fill}
\begin{minipage}[t]{75mm}
\centerline{\includegraphics[scale=0.6]{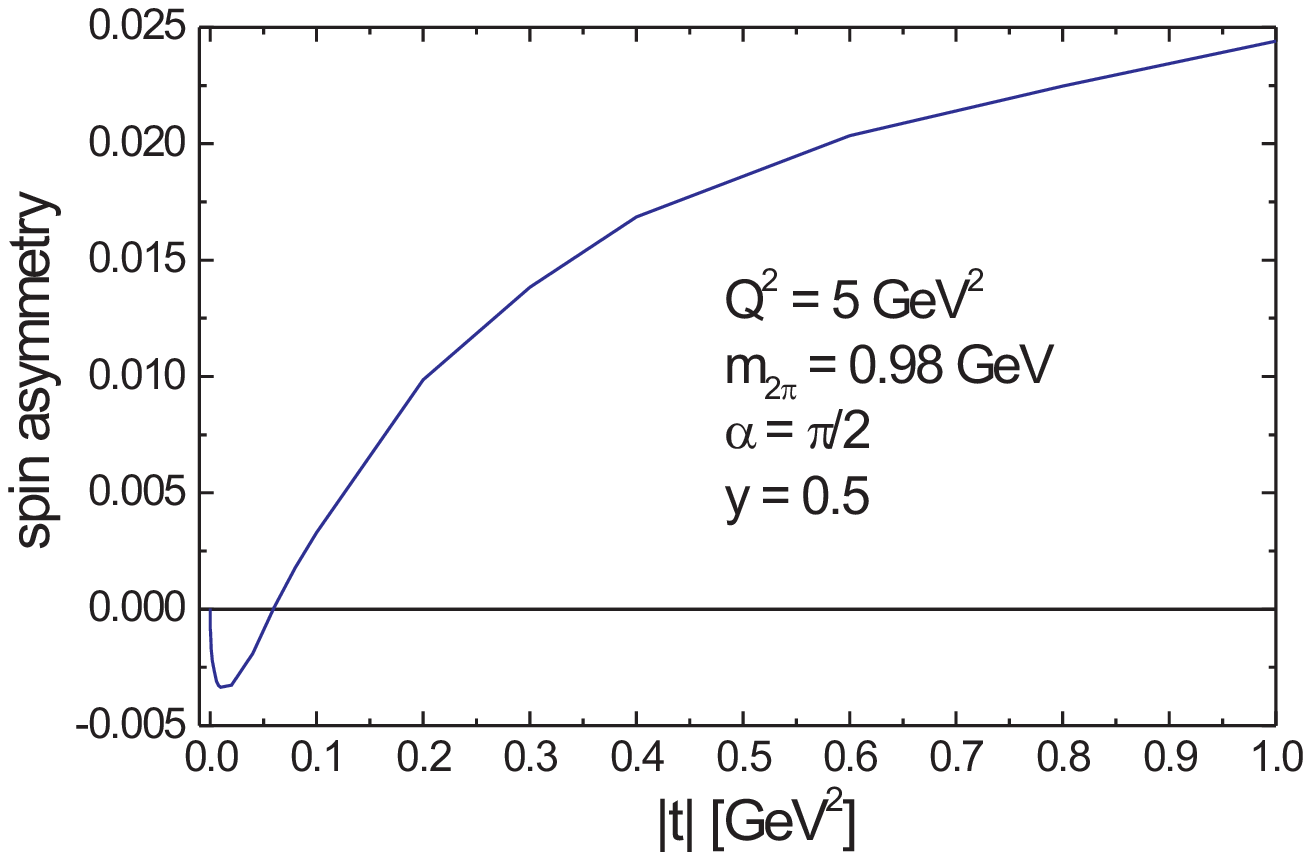}}
\caption{{\protect\small $t-$dependence of the spin asymmetry at
$m_{2\pi}=0.98$
 GeV$^2$, $Q^2=5$ GeV$^2$ }}
\label{tspin}
\end{minipage}
\end{figure}

Here again the asymmetry (\ref{sa}) measures the interference between the
Pomeron and Oddderon exchange amplitudes.
As it is obvious from this equation, the effect is maximal for $\alpha $
near $\pi /2$. The dependence on $m_{2\pi }$ is shown in Fig.~\ref{m2pispin}
for different values of $Q^2$. The  $m_{2\pi }$-dependence is quite
complementary to the
case of charge asymmetry since an additional factor of $i$ comes from the
helicity difference in the leptonic trace, so that the strong phase accumulates
an additional factor of $\pi /2$.
 
Indeed, as the Pomeron amplitude is imaginary
and the Odderon one is real
the relative phase between them is the maximal one for the emergence of
single spin asymmetries \cite{OT01}. The effect should be therefore
maximal for  zero relative phase between isoscalar and isovector
distributions, providing
a complementary probe. Therefore, simultaneous studies
of charge and spin asymmetries provide an important cross-check.

The resulting $m_{2\pi }$ dependence
has thus a characteristic $\cos (\delta_{0,2}-\delta_{1})$ shape,
modulated by the absolute values of the $\rho ,f_{0}$ and $f_{2}$ Breit
Wigner amplitudes. The spin asymmetry is maximal when 
$(\delta_{0,2}-\delta_{1})$
is equal to $0$ or $\pi$, \textit{i.e.} around 0.98 GeV and 1.32 GeV. Let us
also
note
that the $Q^2$-dependence of the spin asymmetry, see Fig. \ref{m2pispin}, is
much weaker than in the case
of charge asymmetry, see Fig. \ref{m2pi}. 
Unfortunately the  magnitude of the
spin asymmetry is quite small (although comparable to
the recent measurement  \cite{Azim}) at low $t$, where
the charge asymmetry is sizeable.

%
%

The $t-$ dependence of the spin asymmetry is shown in Fig.~\ref{tspin}.

%
\section{Sensitivity to the GDA}

The characteristic $m_{2\pi }$ dependence of the asymmetries comes entirely
from the choice
of the two pion distribution amplitude. As we already stressed, this GDA is
a non-perturbative object which we cannot claim to know at present. Our
model was mostly guided by an optimistic expansion of the range of validity
in $m_{2\pi }$ of the Watson theorem up to $1.0$ and even $1.5$ GeV. Other
models \cite{POL} did not use such an assumption, with the important
consequences that neither the drastic phase shift increase near the $f_{0}$
mass \cite{Hyams}, nor the magnitude peak related to it, do appear in the
GDA and therefore in the asymmetries. Because of that we expect that after
taking
into account the $f_{0}$ resonance the estimates of Ref. \cite
{POL} around 1 GeV (where $f_{0}(980)$ contributes) will be modified. On the
other hand, treating the $\pi \pi $ interaction near $1$ GeV without
mentioning the problem of inelasticity and the opening of the $K\bar{K}$
threshold is likely to be unrealistic too. In order to get an estimate of
the effects which may arise due to the opening of the $K\bar{K}$ threshold,
we implemented a modified GDA in our calculation of the charge asymmetry,
which differs from the one given in Eq.~(\ref{GDAI0}) by the inclusion of an
inelasticity factor $\eta (m_{2\pi })$ in front of the $f_{0}$-resonance 
in Eq.~(\ref{GDAI0}) as obtained in the analysis of Ref.~\cite{Kloet}. We show
in
Fig.~(\ref{eta}) the charge asymmetry obtained with such a modified GDA at
$Q^{2}=5$ GeV$^{2}$, $t=-0.8$ GeV$^{2}$ and $\alpha =0$. The effect of this
new parametrization is obviously a decrease of the charge asymmetry above
the $K\bar{K}$ threshold. This change of GDA modifies also moderately the
charge asymmetry in the vicinity of the $f_{2}$-resonance.

One may also adopt an alternative point of view and take these experiments
as another way (together with $\gamma ^{*}\gamma $ reactions \cite{DGP}) to
determine the two pion distribution amplitude, once the dependence of the
asymmetries on variables such as $s$, $Q^{2}$, $t$ and $\alpha $ has been
checked.

\section{Remarks on possible effects of QCD evolution}

The most natural improvement of our results, specially for the charge asymmetry, 
consists in the inclusion of 
the BFKL and BKP evolution in the scattering amplitudes with Pomeron and Odderon
exchanges, $\cal{M}_P$ and $\cal{M}_O$, respectively. This is
beyond the scope of the present
paper, but nevertheless
 we can draw some qualitative conclusions about their possible effects.
There is no $s$-dependence at the Born level, provided $s$ is large enough
for the usual high energy approximation to hold. The BFKL and BKP evolutions
introduce characteristic energy dependences.
They  can also lead to some
changes of the normalization of the involved amplitudes, as well as to the appearence
of some additional phases $\delta_P$ and $\delta_O$.

The charge asymmetry is effectively the product of
$$ 
\frac{| \cal{M}_O |}{|\cal{M}_P|}\cdot \,\sin (\delta_{0,2} - \delta_1)\;.
$$
We believe that the inclusion of BFKL and BKP evolution doesn't change dramatically
the ratio $\frac{| \cal{M}_O |}{|\cal{M}_P|}$. On the other hand, 
the possible appearence of an additional phase difference $\delta = \delta_P - \delta_O$
would lead to a change of the argument of the sine above.
Let us however note that the structure of the charge asymmetry 
with two zeros in Fig.
\ref{m2pi} 
is robust against a moderate (independent of $m_{2\pi}$) phase $\delta$.
The rapid change of the $\delta_0$ phase shift near $m_{2\pi}=1\,$GeV (see Ref.
\cite{Hyams})
enforces a zero of the asymmetry, even if an additional "extra" phase
is introduced.
The same is true for the zero at $m_{2\pi} \approx 0.8\,$GeV. There in
 contrast
to the upper argumentation the rapid change of the pion form factor phase shift
$\delta_1$ enforces the zero.

\begin{figure}[t]
\begin{minipage}[t]{75mm}
\centerline{\includegraphics[scale=0.6]{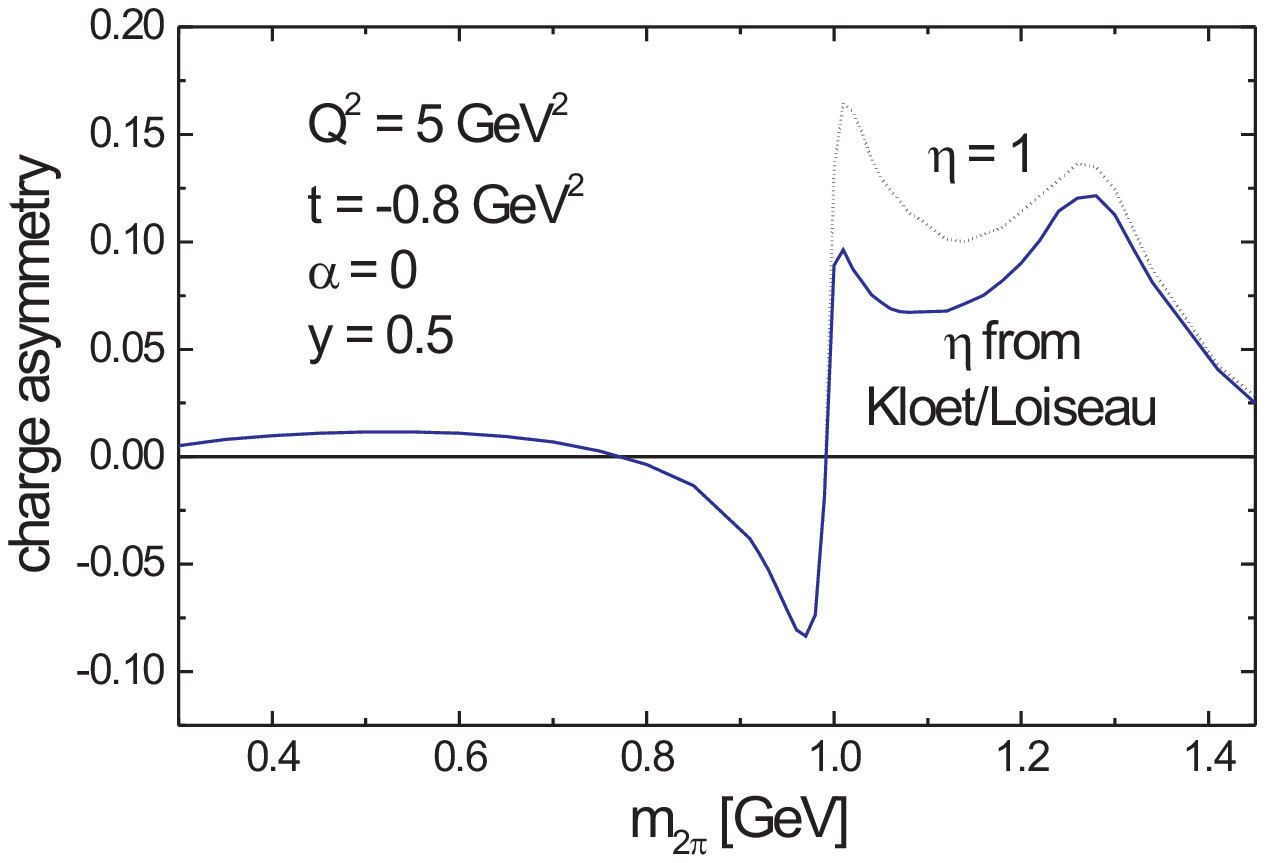}}
\caption{{\protect\small $m_{2\pi}-$dependence of the charge asymmetry
with
the inelasticity factor $\eta (m_{2\pi})$ (solid curve) and with $\eta
(m_{2\pi})=1$ (dotted curve) for $Q^2=5$GeV$^2$, $t=-0.8$GeV$^2$, $%
\alpha=0$, $y=0.5$ }}
\label{eta}
\end{minipage}
\hspace{\fill}
\begin{minipage}[t]{75mm}
\centerline{\includegraphics[scale=0.6]{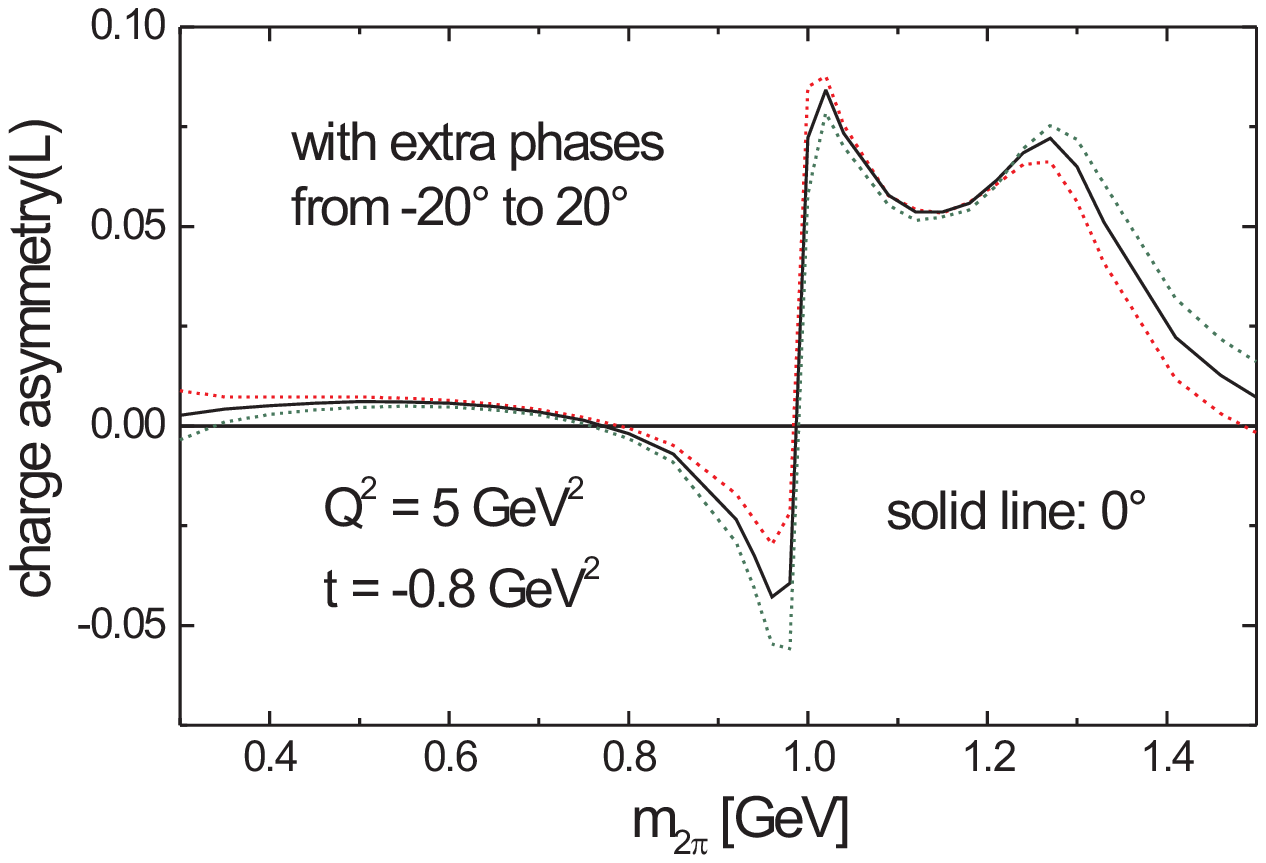}}
\caption{{\protect\small $m_{2\pi}-$dependence of the charge asymmetry from
the longitudinal photon  
with an additional constant phase $\delta$}}
\label{plotphase}
\end{minipage}
\end{figure}

As an illustration of these remarks we 
present in Fig. \ref{plotphase} the longitudinal charge asymmetry
at $Q^2=5\,$GeV$^2$ and $t=-0.8\,$GeV$^2$, calculated
for two values of the additional phase $\delta=\pm 20^\circ$ (dashed lines). 
They resulting curves differ very little from the original curve corresponding 
to $\delta=0^\circ$ (solid line). 

\section{Remarks on HERMES data}

Our discussion here is restricted to diffractive physics, mostly testable in
collider experiments, where center of mass energies are of the order of $100$
GeV and more. At lower energies, exclusive electroproduction of pairs of
mesons are described in the framework of 
the collinear factorization, the soft part of the amplitude being represented by
generalized parton distributions \cite{INPC}. In such a framework 
quark-antiquark and two gluon exchanges play the dominant role.
Here a charge asymmetry may
occur as the result of the interference of charge parity odd and charge
parity even amplitudes. An estimate of this asymmetry has been computed in
Ref. \cite{LDSPG}, using an isosinglet generalized distribution amplitude
which differs from our choice in that it doesn't include the $f_{0}$%
-resonance\footnote{We thank B.~Lehmann-Dronke and M.~Polyakov for discussions on 
this point}. Recent data from the HERMES experiment \cite{HERMES} at HERA are
indeed compatible with these estimates and show confusingly no sign of
the $f_{0}$-resonance.


The estimates in \cite{LDSPG} lead to a very small value of the asymmetries
at low $x_{Bj}$%
. This is easily understandable since in this region gluon exchange diagrams
dominate which select charge parity odd mesonic states, leading to vanishing
interference effects. This opens the interesting possibility that data
unravel an interference effect between two and three gluon exchange, which
would in that framework be understood as a higher twist contribution. In
such a fixed target experiment, a small $x_{Bj}$ value is indeed related
to quite low values of $Q^{2}$ and therefore to higher twist
contributions. One may also understand such an effect as an early sign of
Pomeron-Odderon interference. In any case, pushing the analysis to the
lowest possible $x_{Bj}$ values is extremely interesting.

No single lepton spin asymmetries should show up in these lower energy data
if the leading twist contribution is indeed dominant. The reason is the
asymptotic dominance of the process with a longitudinally polarized virtual
photon. Here also higher twist contributions may yield sizeable spin
asymmetries at quite low values of $Q^2$.




\section{Conclusion}

Our study shows that the role of the Odderon in
diffractive processes in perturbative QCD is intimately related to
a sizeable charge asymmetry in the electroproduction of
two charged mesons. The single-spin asymmetry in the same reaction turned
out to be much
smaller.

We applied the powerfull tool of QCD factorization which allows us to
calculate the hard subprocess perturbatively, while the soft ingredients (GDA
and proton impact factor) should be modeled or, better, measured, which
poses a new challenging problem for experimentalists.

Let us finally emphasize that
data on this reaction in the kinematical domain suitable for our
calculation
(i.e. large $s$, small $t$, $Q^{2}$ above $1\ $GeV$^{2}$ and $m_{2\pi }$
below $1.5\ $GeV) should be 
easily accessible for analysis 
by the experimental
set-ups H1 \cite{Olsson} and ZEUS \cite{ZEUS} at HERA. 
Such confrontation of theory with experiment should shed some   
light on the
status of the Odderon. 

\vskip.1in \textbf{Acknowledgements}

\vskip.1in We acknowledge useful discussions with N.~Bianchi, 
S.~Brodsky, M.~Diehl, H.G.~Dosch, A.V.~Efremov, I.F. Ginzburg, B.~Loiseau,
S.V.~Mikhailov, B.~Nicolescu,
J.~Olsson, T.N.~Pham, A.~Sch{\"a}fer, T.N.~Truong and S.~Wallon.

This work is supported in part by the TMR and IHRP Programmes of the
European Union, Contracts No.~FMRX-CT98-0194 and No.~HPRN-CT-2000-00130.


\begin{thebibliography}{99}

\bibitem{BFKL}  E.A.~Kuraev, L.N.~Lipatov and V.S.~Fadin, Sov. JETP \textbf{%
44} (1976) 443~; \textit{ibid} \textbf{45} (1977) 199~; Ya.Ya.~Balitsky and
L.N.~Lipatov, Sov.\ J.\ Nucl.\ Phys.\ \textbf{28} (1978) 822~; L.N.~Lipatov,
Sov. Phys. JETP \textbf{63} (1986) 904


\bibitem{BKP}  J.~Bartels, Nucl.\ Phys.\ B \textbf{151} (1979) 293; \textit{%
ibid} B \textbf{175} (1980) 365; \newline
J.~Kwiecinski and M.~Praszalowicz, 
Phys.\ Lett.\ B \textbf{94} (1980) 413. 

\bibitem{Levodd}  L.N. Lipatov, Phys.\  Lett.\ B \textbf{309} (1993) 394~;
JETP Lett. \textbf{59} (1994) 596~; Sov.\ Phys.\ JETP\ Lett.\ \textbf{59}
(1994) 571; \newline
L.D.~Fadeev and G.P. Korchemsky, Phys.\ Lett.\ B \textbf{342} (1995) 311

\bibitem{JW}  R.A.~Janik and J.~Wosiek, Phys. Rev. Lett. \textbf{82} (1999)
1092

\bibitem{Vacca1}  J.~Bartels, L.N.~Lipatov and G.P.~Vacca, Phys. Lett. B
\textbf{477} (2000) 178

\bibitem{Korch}  G.~P.~Korchemsky, J.~Kotanski and A.~N.~Manashov,
hep-ph/0111185. 



\bibitem{LN}  L.~Lukaszuk and B.~Nicolescu, Lett. Nuovo Cim. \textbf{8}
(1973) 405


\bibitem{Doshrecent}  H.G.~Dosch, C.~Ewerz and V.~Schatz, hep-ph/0201294



\bibitem{Olsson}  J.~Olsson (for the H1 Collab.), hep-ex/0112012

\bibitem{Dosh}  E.R.~Berger, A.~Donnachie, H.G.~Dosch, W.~Kilian,
O.~Nachtmann and M.~Reuter, Eur. Phys. J. C \textbf{9} (1999) 491



\bibitem{KM}  J.~Czyzewski, J.~Kwiecinski, L.~Motyka and M.~Sadzikowski,
Phys. Lett. B \textbf{398} (1997) 400~; erratum \textit{ibid} B \textbf{411}
(1997) 402

\bibitem{Engel}  R.~Engel, D.Yu.~Ivanov, R.~Kirschner and L.~Szymanowski,
Eur. Phys. J. C \textbf{4} (1998) 93

\bibitem{Vacca2}  J.~Bartels, M.A.~Braun, D.~Colferai and G.P.~Vacca, Eur.
Phys. J. C \textbf{20} (2001) 323



\bibitem{Brodsky}  S.J.~Brodsky, J.~Rathsman and C.~Merino, Phys. Lett. B
\textbf{461} (1999) 114

\bibitem{Nikolaev}  I.P.~Ivanov, N.N.~Nikolaev and I.F. Ginzburg,
hep-ph/0110181


\bibitem{HPST}  P.~H\"{a}gler, B.~Pire, L.~Szymanowski and O.~V.~Teryaev,
Phys. Lett. \textbf{B 535} (2002) 117, hep-ph/0202231, and erratum to be
published, \\
P.~H\"{a}gler, B.~Pire, L.~Szymanowski and O.~V.~Teryaev,
hep-ph/0206270.

\bibitem{DGPT}  M.~Diehl, T.~Gousset, B.~Pire and O.V.~Teryaev,
Phys.\ Rev.\ Lett.\ \textbf{81} (1998) 1782 



\bibitem{IF}
I.~F.~Ginzburg, S.~L.~Panfil and V.~G.~Serbo,
Nucl.\ Phys.\ B {\bf 284} (1987) 685;
{\it ibid}
Nucl.\ Phys.\ B {\bf 296} (1988) 569.


\bibitem{GI}  I.F.~Ginzburg and D.Yu.~Ivanov, Nucl. Phys. B \textbf{388}
(1992) 376

\bibitem{Azim}
A.~Airapetian {\it et al.}  [HERMES Collaboration],
Phys.\ Rev.\ Lett.\  {\bf 84}, 4047 (2000);\\
H. Avakian, talk presented at the Baryons 2002 Conference


\bibitem{OT01}
O.V. Teryaev, {\it T-odd diffractive distributions},
in Proceedings of IX Blois Workshop, p.211-216.

\bibitem{POL}  M.~V.~Polyakov and C.~Weiss,
Phys.\ Rev.\ D \textbf{59} (1999) 091502 [arXiv:hep-ph/9806390];

M.~V.~Polyakov, Nucl.\ Phys.\ B \textbf{555} (1999) 231

\bibitem{DGP}  M.~Diehl, T.~Gousset and B.~Pire,
Phys.\ Rev.\ D \textbf{62} (2000) 073014 

\bibitem{protonP}  J.F.~Gunion and D.E.~Soper, Phys. Rev. D \textbf{15}
(1977) 2617

\bibitem{protonO}  M.~Fukugita and J.~Kwiecinski, Phys. Lett. B \textbf{83}
(1979) 119

\bibitem{Hyams}  B.~Hyams \textit{et al.}, Nucl.\ Phys.\ B \textbf{64}
(1973) 134 \newline
D.V.~Bugg, B.S.~Zou and A.V.~Sarantsev, Nucl.\ Phys.\ B\textbf{471} (1996)
59 \newline
R.~Kaminski, L.~Lesniak and K.~Rybicki, Acta Phys.Polon. B \textbf{31}
(2000) 895

\bibitem{LDSPG}  B.~Lehmann-Dronke, P.~V.~Pobylitsa, M.~V.~Polyakov, A.~Sch{%
\"{a}}fer and K.~Goeke,
Phys.\ Lett.\ B \textbf{475} (2000) 147 [arXiv:hep-ph/9910310];

B.~Lehmann-Dronke, A.~Sch{\"{a}}fer, M.~V.~Polyakov and K.~Goeke,
Phys.\ Rev.\ D \textbf{63} (2001) 114001 [arXiv:hep-ph/0012108].

\bibitem{INPC}  T. Gousset, M. Diehl, B. Pire and O. Teryaev, Nucl. Phys.
\textbf{A654} (1999) 576c

\bibitem{HERMES}   N.~Bianchi private communication.


\bibitem{Kloet}  
W.~M.~Kloet and B.~Loiseau,
Z.\ Phys.\ A \textbf{353} (1995) 227 [arXiv:nucl-th/9408025].

\bibitem{ZEUS}   
J.~Breitweg \textit{et al.} [ZEUS Collaboration],
Eur.\ Phys.\ J.\ C \textbf{6} (1999) 603 [arXiv:hep-ex/9808020].


\end{thebibliography}
\end{document}